\documentclass[11pt]{article}
\usepackage{times}
\usepackage{geometry}
\usepackage[utf8]{inputenc}
\usepackage{enumitem,amssymb}
\usepackage{ragged2e}
\newlist{thematic}{itemize}{8}
\setlist[thematic]{label=$\square$}
\usepackage{pifont}
\usepackage{fancyhdr}
\usepackage[USenglish]{babel}
\usepackage[utf8]{inputenc}
\usepackage[T1]{fontenc}
\usepackage{epsfig}
\usepackage{wrapfig}
\usepackage{color}

\usepackage[vflt]{floatflt}
\usepackage{longtable}
\usepackage{caption}
\usepackage[dvipsnames,svgnames,x11names]{xcolor}
\usepackage[colorlinks=true, linktocpage, linkcolor={blue!40!black}, citecolor={blue!40!black}, urlcolor={blue!50!black}]{hyperref}
\usepackage[all]{hypcap} 
\usepackage{amssymb}
\usepackage{amsmath, xspace}
\usepackage{graphics,graphicx} 
\usepackage{rotating}
\usepackage{color}
\usepackage{xcolor}
\usepackage{multirow}
\usepackage{subfigure}
\usepackage{sectsty}
\usepackage[outercaption]{sidecap}
\sidecaptionvpos{figure}{c}

\usepackage[authoryear,round]{natbib}
\usepackage[authoryear,round]{natbib}
\usepackage{bibentry}
\nobibliography*   
\definecolor{DarkGreen}{rgb}{0.0, 0.3, 0.0}
\definecolor{purple}{rgb}{0.5, 0.0, 0.5}
\definecolor{red}{rgb}{1, 0.0, 0.0}
\definecolor{green}{rgb}{0, 1.0, 0.0}

\newcommand{\Moon}{$M_{Moon}$}                          

\newcommand{\apjl}{\rm ApJL}
\newcommand{\apjs}{\rm ApJS}

\newcommand{\aap}{\rm A$\&$A}

\def\ao{{\it Appl.~Opt.}}           

\def\3he{$^3{\rm He}$}
\def\arcdeg{\hbox{$^\circ$}}

\def\lsim{\mathrel{\lower2.5pt\vbox{\lineskip=0pt\baselineskip=0pt
           \hbox{$<$}\hbox{$\sim$}}}}

\def\gsim{\mathrel{\lower2.5pt\vbox{\lineskip=0pt\baselineskip=0pt
           \hbox{$>$}\hbox{$\sim$}}}}

\begin{document}
\raggedright
\Large {\bf
The search for dark matter axions from neutron stars in the inner parsecs of the Milky Way with AtLAST}
\linebreak
\bigskip
\normalsize


{\bf
J. De Miguel$^{1, 2, 3}$ (jdemiguel@iac.es),
E. Hatziminaoglou$^{4, 1, 3}$, 
J. Prieto-Polo$^3$, 
J. D. Marrero-Falcón$^3$, Abaz Kryemadhi$^4$
}

{\it 
$^1$ Instituto de Astrof\'isica de Canarias (IAC), E-38200 La Laguna, Tenerife, Spain;\\ 
$^2$ The Institute of Physical and Chemical Research (RIKEN), 
Center for Advanced Photonics, 519-1399 Aramaki-Aoba, Aoba-ku, Sendai, Miyagi 980-0845, Japan;\\
$^3$ Departamento de Astrof\'isica, Universidad de La Laguna, E-38206 La Laguna, Tenerife, Spain\\
$^4$ ESO, Karl-Schwarzschild-Str. 2, 85748 Garching bei M{\"u}nchen, Germany 

$^4$ Dept. Computing, Math, and Physics, Messiah University, Mechanicsburg, PA, US }

\vskip 0.3cm
\textbf{Science Keywords:} 
cosmology: dark matter; stars: pulsars, magnetars; Galaxy: center.
\linebreak

 \captionsetup{labelformat=empty}
\begin{figure}[h]
   \centering
\includegraphics[width=.9\textwidth]{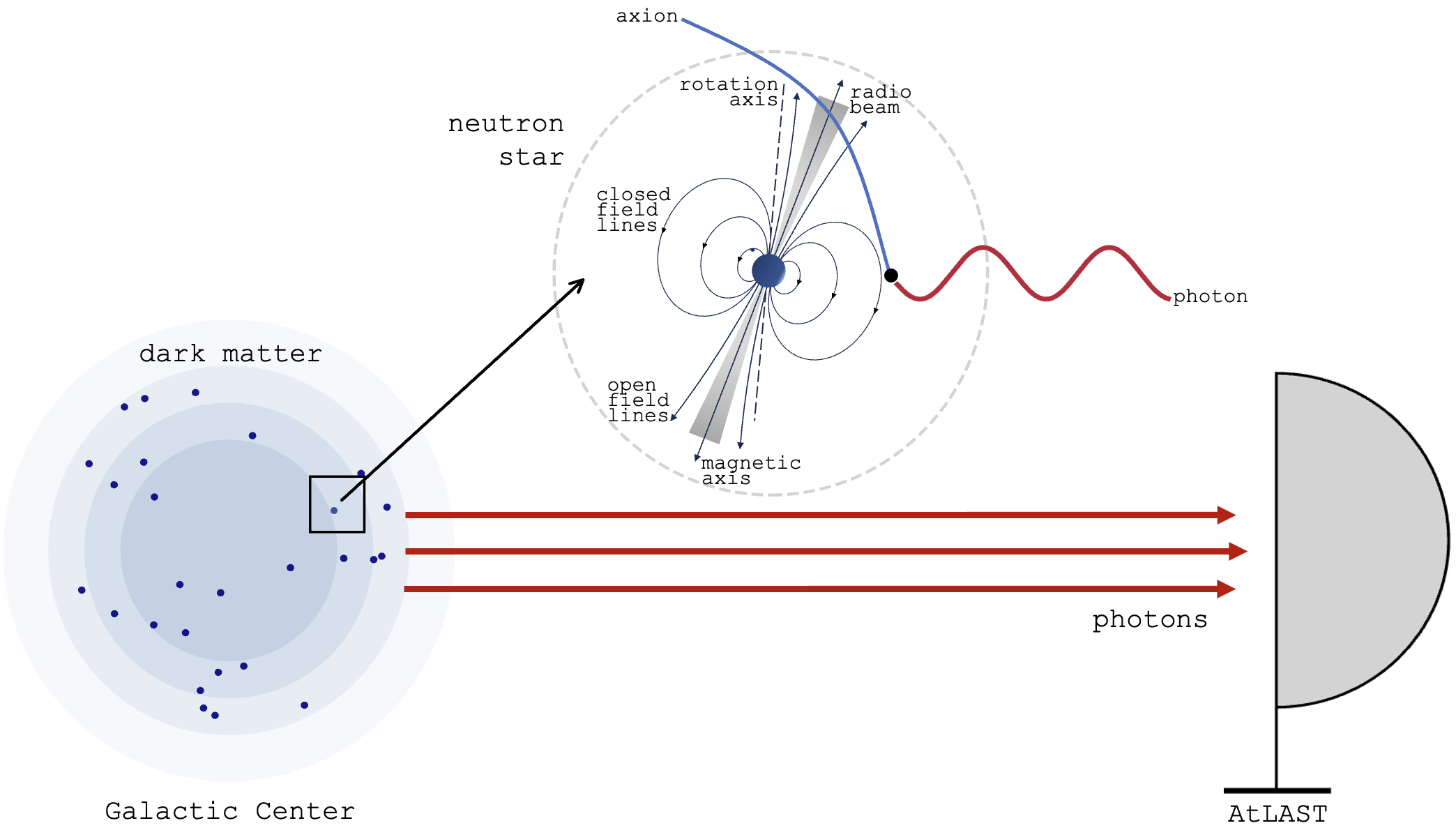}
   \caption{Conceptual overview of the proposed AtLAST search for Dark Matter axions in the Galactic Centre.}
\end{figure}

\setcounter{figure}{0}
\captionsetup{labelformat=default}


\begin{justify}
\section*{Abstract}
The magnetospheres of neutron stars in the Galactic Center provide an exceptional environment to search for dark matter axions through their resonant conversion into photons. The combination of extreme magnetic fields and high dark matter density in this region creates ideal conditions for axion–photon conversion across the centimetre–to–submillimetre wavelength range. Detecting the cumulative emission from these neutron star populations demands a facility with unprecedented sensitivity, spatial resolution, and broad frequency coverage in the submillimetre domain, capabilities essential to explore a previously inaccessible region of axion parameter space.
\pagebreak
\section{Scientific context and motivation}

The axion \citep{PhysRevLett.40.223, PhysRevLett.40.279}, a pseudoscalar boson proposed to solve the strong CP problem in quantum chromodynamics, is a compelling dark matter (DM) candidate \citep{1933AcHPh...6..110Z, 1970ApJ...159..379R, ABBOTT1983133, DINE1983137, PRESKILL1983127}. 
Axion mass is predicted by theory and constrained by cosmology and experiments, see e.g., \citet{PhysRevD.109.062002, sym16020163, Hernandez-Cabrera:2023syh, PhysRevD.111.023016, DeMiguel:2020rpn} and references therein. 

In a magnetic field ($B$), axions ($a$) convert into photons ($\gamma$) through $a + B \leftrightarrow \gamma$. Neutron stars (NSs), with magnetic fields of $\sim$$10^{12}$–$10^{15}$ G, provide extreme conditions for axion–photon conversion \citep{PhysRevLett.121.241102, Safdi:2018oeu}. In NS magnetospheres in particular, ambient axions resonantly convert at the local plasma frequency, producing narrow spectral lines. Several radio surveys (e.g. VLA, MeerKAT) have searched for such features between 1–40 GHz ($10^{-5}\!\lesssim\! m_a\!\lesssim\!10^{-4}$ eV) without detecting any axion-compatible signature \citep{Darling:2020plz, Darling:2020uyo, PhysRevLett.125.171301, Foster:2022fxn}. The resulting exclusion limits are shown in Fig.~\ref{Fig_2}.

\begin{figure*}[h]\centering
\includegraphics[width=.8\textwidth]{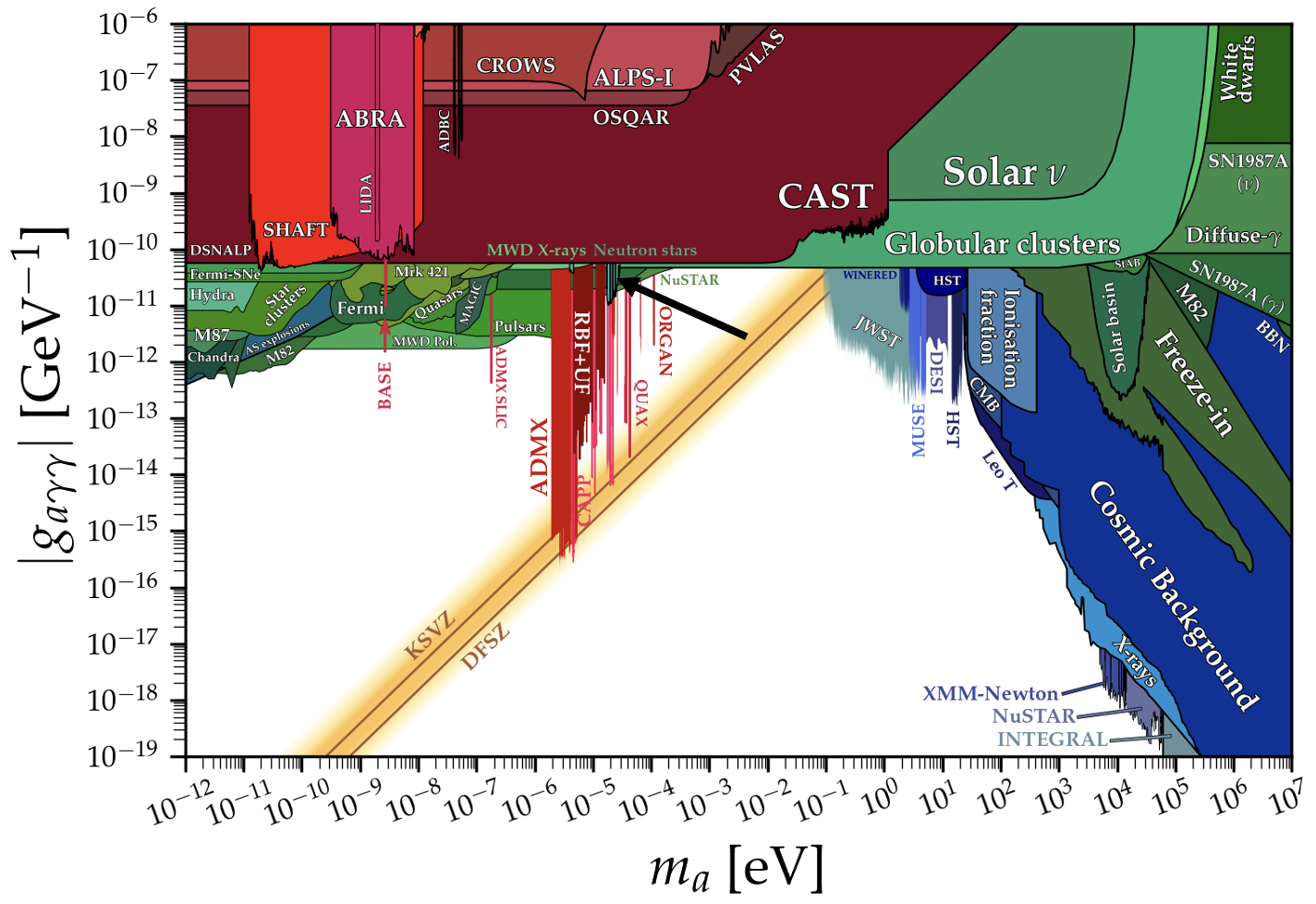}
\caption{
Current limits on the axion--photon coupling, $g_{a\gamma\gamma}$, as a function of axion mass. The green band labelled ``Neutron stars'' is relevant to this white paper (black arrow). Reprinted from \citet{ciaran_o_hare_2020_3932430}.
}
\label{Fig_2}
\end{figure*}
In this white-paper we show that the search for axion-induced signatures from the NS population within the inner parsec of the Milky Way (MW) can be optimally carried out with a facility possessing the characteristics of the Atacama Large Aperture Submillimeter Telescope (AtLAST), thanks to its large collecting area, high sensitivity, and direct access to the Galactic Center (GC).


\vspace{-3mm}
\section{Science case}

The current picture of pulsar-type NSs involves pair cascades and particle acceleration in their magnetospheres. Charges reach relativistic energies, producing curvature, synchrotron, and inverse-Compton emission. Pair cascades are favoured in regions where the electric field is weakly screened, allowing efficient acceleration that generates $\gamma$-rays and sustains a quasi-periodic feedback loop, including the polar cap near the magnetic poles, the outer gap toward the light cylinder, and the slot gap \citep{1975ApJ...196...51R, 1979ApJ...231..854A, 1997A&A...322..846K}. Therefore, magnetospheric splits may deviate substantially from the classical Goldreich \& Julian (GJ) model, which assumes stationary, corotating charges \citep{1969ApJ...157..869G}. In extremely magnetic NSs, referred to as magnetars, such deviations are expected to be considerably stronger, possibly involving most of the magnetosphere \citep{Beloborodov_2012, Beloborodov_2013}. 

The axion-induced flux from an isolated NS scales as
\begin{equation}
S_\nu \propto f(t)\, d^{-2}\, g_{a\gamma\gamma}^{\,2}\, R_*\, m_a^{4/3}\, B_0^{1/3}\, \Omega_*^{-8/3}\, \rho_a\, M_*\, v_0^{-1}\, v_a\,\Big(\tfrac{\mathcal{M}}{\gamma_p}\Big)^{-3/2}
\label{eq:scaling}
\end{equation}
In Eq. \ref{eq:scaling}, the temporal and geometrical modulation is described by $f(t)$, $m_a$ is the axion mass, $g_{a\gamma\gamma}$ the axion–photon coupling, $d$ the source distance, $R_*$ the star radius, $B_0$ the surface magnetic field, $\Omega_*$ the spin frequency, $\rho_a$ the axion density, $M_*$ the star mass, $v_0$ the star velocity and $v_a$ the axion velocity. \cite{DEMIGUEL2025139328}, \cite{DeMiguel:2025rmh}, and \cite{PhysRevD.106.L041302, DeMiguel:2021pfe} showed that the plasma pair multiplicity ($\mathcal{M}$) and Lorentz factor ($\gamma_p$) can shift the axion–photon resonance to higher frequencies, namely from a few GHz up to the mm band, pushing the detectable spectral features in a poorly explored axion parameter space. 

The GC offers an exceptionally favorable environment to search for axion with NSs. As shown by \citet{McMillan_2016} and \citet{2018A&A...619A..46L}, the central parsec may reach DM densities as high as $\rho_{\mathrm{DM}}\!\sim\!10^8$\,GeV\,cm$^{-3}$, up to $10^9$ times denser than the local halo. Furthermore, as noted by \citet{Safdi:2018oeu} and \citet{Foster:2022fxn}, a population of $n$ NSs along the line-of-sight would boost the axion-induced integrated signal. 

The expected emission from the inner parsec is dominated by relatively young to middle-aged pulsars (ages $\lesssim$1\,Myr), which exhibit high pair multiplicities ($\mathcal{M}\lesssim10^{5}$; \citealt{Timokhin:2015dua, Timokhin:2018vdn}). Using population estimates from \citet{Foster:2022fxn}, about $n\!\sim\!10^5$ NSs may reside within the central region of the MW. While the signal from a single pulsar in the GC would likely be too faint to detect individually, the combined emission from the entire population, amplified by the DM spike, could render axion detection feasible. The expected axionic signal from young pulsars could extend up to, roughly, 1 cm wavelength. Given their stronger magnetization and higher plasma frequencies, magnetars constitute ideal targets for axion searches extending into the sub-mm band, as illustrated in Fig.~\ref{Fig_0}. Among the few dozen magnetars known, one (SGR~1745$-$2900) lies in the GC \citep{Mori:2013yda}. The Galactic magnetar birth rate is estimated at $1.8^{+0.6}_{-2.6}$\,kyr$^{-1}$, implying that about $10.7^{+4.4}_{-18.8}\%$ of all NSs are born as magnetars \citep{Sautron_2025}.
\begin{figure*}[h]\centering
\includegraphics[width=.63\textwidth]{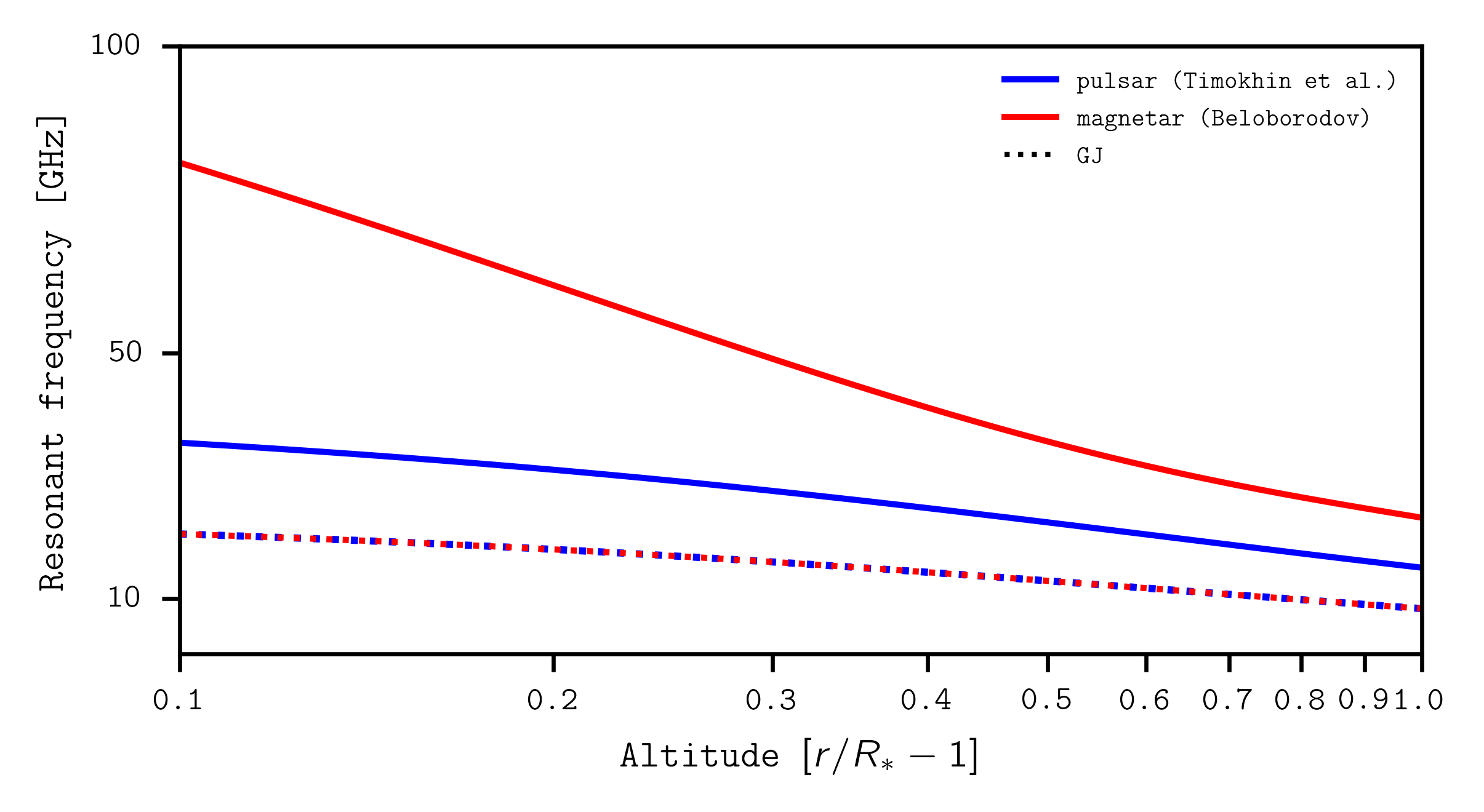}
\caption{Resonant axion--photon conversion frequency versus altitude above the NS surface. The calculation \citep{DEMIGUEL2025139328} compares two magnetospheric models: a young pulsar (blue; Timokhin et~al.) and a magnetar (red; Beloborodov). Solid lines mark the resonance condition $\omega_p\!\sim\!m_a$, and dotted lines the GJ prediction. Pulsar: $B_0=10^{13}\,$G, $P=0.1$\,s; magnetar: $B_0=5\times10^{14}$\,G, $P=5$\,s. All calculations assume $R_\ast=12$\,km. }
\label{Fig_0}
\end{figure*}

Current single-dish facilities can point to the GC (Dec $\simeq -29^\circ$), but only the LMT in Mexico (lat $\simeq 19^\circ$N) reaches a favorable elevation of $\sim42^\circ$ at culmination. 
IRAM~30\,m ($\sim24^\circ$), GBT ($\sim23^\circ$), and Yebes~40\,m ($\sim20.5^\circ$) achieve only marginal elevations, while Effelsberg ($\sim10.5^\circ$) is effectively impractical. 
AtLAST’s 50\,m aperture operates without confusion limitations in the cm-to-mm regime and provides beams narrower than the complex GC background.

\section{Technical requirements}

AtLAST uniquely combines a large collecting area, a surface accuracy of $\sim$15\,$\mu$m enabling efficient operation in the sub-mm domain, a $2^\circ$ instantaneous field-of-view (FoV), and a southern location with excellent access to the GC. These characteristics together provide a level of sensitivity and survey speed unattainable by any existing single-dish or interferometric facility.

At a distance of 8.3\,kpc, a radius of 1\,pc subtends about $25''$, corresponding to a diameter of $\sim50''$ (0.83$'$) on the sky. The diffraction-limited beam of a 50\,m telescope is given by HPBW $\simeq 1.2\,\lambda/D$, yielding $\sim50''$ at 30\,GHz ($\lambda \approx 1$\,cm), which matches the angular size of the region and allows the entire central parsec to be observed in a single pointing. At 150\,GHz ($\lambda \approx 2$\,mm), the beam shrinks to $\sim10''$, so mapping the same region would require a small mosaic of about 25 beams. Thanks to AtLAST's wide FoV and rapid mapping capability, such coverage is technically straightforward.

Lastly, a spectral resolution of a few MHz is sufficient for resolving the narrow axion-induced features expected from resonant photon conversion. The corresponding full width at half maximum scales as
\begin{equation}
\Delta\nu/\nu \sim \Omega_* r \varepsilon^2/c, \,
\end{equation}
with $r$ being the conversion altitude, $\varepsilon$ the eccentricity for an oblique rotator at which the intersection of a plane
perpendicular to the rotation axis with the conversion surface
is projected an ellipse, and $c$ the speed of light. This full width at half maximum, of the order of tens of MHz, is well matched to the spectral resolution achievable with AtLAST’s instrumentation.

\end{justify}


\noindent 
{\bf References: } Abbott L. \& Sikivie P. 1983, Phys. Lett. B, 120, 133 $\bullet$ 
Arons J. \& Scharlemann E. T. 1979, ApJ, 231, 854 $\bullet$ 
Beloborodov A. M. 2012, ApJ, 762, 13 $\bullet$ 
Beloborodov A. M. 2013, ApJ, 777, 114 $\bullet$ 
Darling J. 2020a, Phys. Rev. Lett., 125, 121103 $\bullet$ 
Darling J. 2020b, ApJL, 900, L28 $\bullet$ 
De Miguel J. et al. 2025a, arXiv:2512.06441 [hep-ph] $\bullet$
De Miguel J. 2025, Phys. Lett. B, 862, 139328 $\bullet$ 
De Miguel J. 2021, JCAP, 04, 075 $\bullet$
De Miguel J. 2025, Phys. Lett. B, 862, 139328 $\bullet$ 
De Miguel J. et al. 2024, Phys. Rev. D, 109, 062002 $\bullet$ 
De Miguel J., Kryemadhi A. \& Zioutas K. 2025, Phys. Rev. D, 111, 023016 $\bullet$ 
De Miguel J., \& Otani C. 2022a, Phys. Rev. D, 106, L041302 $\bullet$ 
De Miguel J., \& Otani C. 2022b, JCAP, 08, 026 $\bullet$ 
Dine M. \& Fischler W. 1983, Phys. Lett. B, 120, 137 $\bullet$ 
Foster, J. W. et al. 2022, Phys. Rev. Lett., 129, 251102 $\bullet$ 
Foster J. W. et al. Phys. Rev. Lett., 125, 171301 $\bullet$
Goldreich P. \& Julian W. H. 1969, ApJ, 157, 869 $\bullet$ 
Hernández-Cabrera J. F. et al. 2024a, JINST, 19, P01022 $\bullet$
Hernández-Cabrera J. F. et al. 2024b, Symmetry 16 $\bullet$
Hook A. et al. Phys. Rev. Lett., 121, 241102 $\bullet$ 
Kramer M. et al. 1997, A\&A, 322, 846 $\bullet$ 
Lacroix T. 2018, A\&A, 619, A46 $\bullet$ 
McMillan P. J. 2016, MNRAS, 465, 76 $\bullet$ 
Mori K. et al. 2013, ApJL, 770, L23 $\bullet$
O’Hare, C. 2020, cajohare/AxionLimits: AxionLimits, v1.0 $\bullet$ 
Preskill J., Wise M. B. \& Wilczek, F. 1983, Phys. Lett. B, 120, 127 $\bullet$ 
Rubin V. C. \& Ford W. Kent, J. 1970, ApJ, 159, 379 $\bullet$ 
Ruderman M. A. \& Sutherland P. G. 1975, ApJ, 196, 51 $\bullet$ 
Safdi B. R., Sun Z. \& Chen A. Y. 2019, Phys. Rev. D, 99, 123021 $\bullet$
Sautron M. et al. 2025, ApJ, 986, 88 $\bullet$ 
Timokhin A. N. \& Harding A. K. 2015, ApJ, 810, 144 $\bullet$ 
Timokhin A. N. \& Harding A. K. 2019, ApJ, 871, 12 $\bullet$
Weinberg S. 1978, Phys. Rev. Lett., 40, 223  $\bullet$
Wilczek F. 1978, Phys. Rev. Lett., 40, 279 $\bullet$ 
Zwicky, F. 1933, Helvetica Physica Acta, 6, 110.

\nobibliography{apssamp}

\end{document}